\documentclass[aps,prl,10pt,twocolumn,superscriptaddress]{revtex4-1}  

\usepackage{xcolor}
\usepackage{ulem}

\usepackage{graphicx}  
\usepackage{dcolumn}   
\usepackage{bm}        
\usepackage[utf8]{inputenc} 
\usepackage[T1]{fontenc} 
\usepackage{amsmath,amssymb,amsthm,stmaryrd} 
\usepackage{epstopdf}  
\usepackage{graphics}       
\usepackage[english]{babel} 
\usepackage{listing}  

\newcommand{\N}{\mathbb N}

\newcommand\Pe{{\rm Pe}} 

\DeclareMathSymbol{\leqslant}{\mathalpha}{AMSa}{"36} 
\DeclareMathSymbol{\geqslant}{\mathalpha}{AMSa}{"3E} 
\DeclareMathSymbol{\eset}{\mathalpha}{AMSb}{"3F}     
\renewcommand{\leq}{\;\leqslant\;}                   
\renewcommand{\geq}{\;\geqslant\;}                   

\newcommand{\bra}{\langle}
\newcommand{\ket}{\rangle}

\newcommand{\ga}{\alpha}
\newcommand{\gb}{\beta}
\newcommand{\gga}{\gamma}            
\newcommand{\gd}{\delta}

\newcommand{\gr}{\rho}

\newcommand{\gD}{\Delta}

\newcommand{\gl}{\lambda}

\newcommand{\gs}{\sigma}

\renewcommand{\cosh}{\,{\rm ch}}
\renewcommand{\sinh}{\,{\rm sh}}
\usepackage{enumitem}
\setlist[itemize]{leftmargin=4mm,topsep=.25cm,parsep=.1cm,itemsep=0cm,label=$\bullet$}
\setlist[enumerate]{leftmargin=6mm,topsep=0.05cm,parsep=.1cm,itemsep=0cm}
\usepackage[pdftex]{hyperref}
\begin{document}

\title{Exact Hydrodynamic Description of Active Lattice Gases}

\author{Mourtaza Kourbane-Houssene}
\affiliation{Universit\'e Paris Diderot, Sorbonne Paris Cit\'e, MSC, UMR 7057 CNRS, 75205 Paris, France}
\author{Cl\'ement Erignoux} \affiliation{Instituto de Matem\'atica Pura e Aplicada, Rio de Janeiro, Brazil}
\author{Thierry Bodineau} \affiliation{
CMAP, Ecole polytechnique, CNRS, Universit\'e Paris-Saclay, 91128, Palaiseau, France}
\author{Julien Tailleur}
\affiliation{Universit\'e Paris Diderot, Sorbonne Paris Cit\'e, MSC, UMR 7057 CNRS, 75205 Paris, France}

\date{\today}

\begin{abstract}
We introduce a class of lattice gas models of active matter systems
whose hydrodynamic description can be derived exactly. We illustrate
our approach by considering two systems exhibiting two of the most
studied collective behaviours in active matter: the motility-induced
phase separation and the transition to collective motion. In both
cases, we derive coupled partial differential equation describing the
dynamics of the local density and polarization fields and show how
they quantitatively predict the emerging properties of the macroscopic
lattice gases.
\end{abstract}

\maketitle




Active matter systems are intrinsically out of thermal equilibrium due
to the dissipation of energy at the microscopic scales to produce
motion~\cite{Marchetti:2013:RMP,Romanczuk:2012:EPJST,Vicsek:2012:PhysRep,Bechinger:2016:RMP}. The
resulting non-Brownian random walks endow these systems with a rich
phenomenology, from the long-range order observed in 2D assemblies of
self-propelled particles~\cite{Vicsek:1995:PRL,Toner:1995:PRL,Schaller:2010:Nature,Bricard:2013:Nature} to the
spatio-temporal chaos of dense assemblies of nematic
particles~\cite{Wensink:2012:PNAS,Ngo:2014:PRL} through the enhanced clustering
resulting from the interplay of repulsive forces and
self-propulsion~\cite{Cates:2015:ARCMP,Theurkauff:2012:PRL,Palacci:2013:Science}.

The toolbox of equilibrium statistical mechanics cannot be used
\textit{a priori} to describe such non-thermal systems and one has to
rely on dynamical studies, even to characterize systems in a steady
state. When an effective detailed-balance with respect to a
non-Boltzman distribution is (partially)
restored~\cite{Fodor:2016:PRL,Tailleur:2008:PRL,Tailleur:2009:EPL},
this can only be established after a complex, case-by-case study of
otherwise analytically untractable dynamics. Numerical simulations
have thus become a prominent tool to study active matter, and progress
is often hindered by strong finite-size
effects~\cite{Gregoire:2004:PRL}. In such contexts, exact results
derived on simple model systems can offer much needed guiding
principles. Whereas this has frequently been true outside equilibrium,
for instance to characterize dynamical phase
transitions~\cite{Derrida:2007:JSM}, little success has been achieved
along these lines for active matter systems. In particular, the
derivation of coarse-grained descriptions of active systems has
attracted a lot of interest over the past decades~\cite{Bertin:2006:PRE,Ahmadi:2006:PRE,Bertin:2009:JPA,Baskaran:2009:PNAS,Baskaran:2010:JSM,Ihle:2011:PRE,Peshkov:2012:PRL,Thuroff:2013:PRL,Solon:2013:PRL,Bertin:2013:NJP,Ihle:2014:EPJST,Bertin:2015:PRE,Manacorda:2017:PRL}, but the
complexity of the underlying microscopic models has prevented the
derivation of exact results outside the mean-field
regime~\cite{Bolley:2012:AML,Degond:2008:MMMAS}.

In this letter, following the recent interest for lattice models of
active
particles~\cite{Czirok:1997:JPA,Oloan:1999:JPA,Peruani:2011:PRL,Thompson:2011:JSM,Solon:2013:PRL,Soto:2014:PRE,Pilkiewicz:2014:PRE,Manacorda:2017:PRL},
we introduce a new class of active lattice gas models whose
large-scale physics can be described exactly. We build on recent
developments in the mathematical-physics literature to derive exact
hydrodynamic descriptions of lattice
gases~\cite{DeMasi:1985:PRL,JonaLasinio:Proba:1993, Bodineau:2010:JSP,
  Erignoux:2016:Arxiv}. For illustration purposes, we focus on two of
the most studied emergent behaviours in active systems: the
motility-induced phase { separation (MIPS)}
~\cite{Tailleur:2008:PRL,Fily:2012:PRL,Cates:2015:ARCMP} and the
transition to collective
motion~\cite{Vicsek:1995:PRL,Gregoire:2004:PRL,Solon:2015:PRL}, but
the approach we present here can be extended beyond these cases. For
both systems, we single out the relevant hydrodynamic modes and
construct their exact dynamics. This allows us both to simulate
efficiently their large scale behaviours as well as analytically study
their instabilities and the corresponding phase diagrams.

\begin{figure}
  $\begin{array}{cc}
    \hspace{-0.5cm} \includegraphics[scale=0.9]{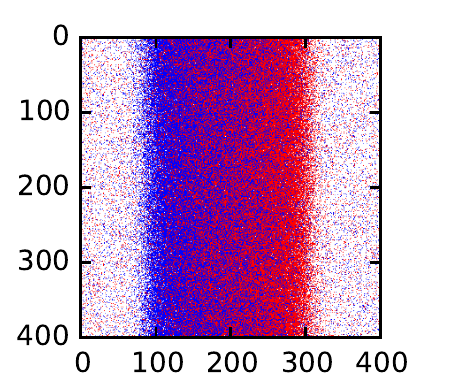}
    &\hspace{-0.5cm}\includegraphics[scale=0.9]{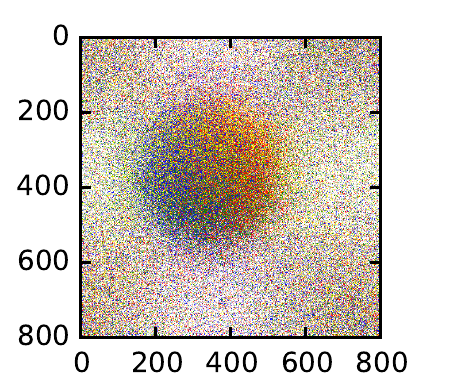}
  \end{array}$
  \caption{Snapshots of microscopic simulations of dynamics 1.1-1.3 in
    2D showing MIPS for lattices of $\alpha L\times \alpha L$ sites,
    with $L=100$. The symmetry of the dense phase depends whether
    particles are biased only along $\hat x$ (left,
    $\alpha=4,\,\rho_0=0.65$) or can point along the four lattice
    bonds (right, $\alpha=8,\,\rho_0=0.63$). The color of a site gives
    the direction of a particle: blue for $\shortrightarrow$, green
    for $\shortuparrow$, red for $\shortleftarrow$, yellow for
    $\shortdownarrow$.  Simulations
    parameters: $D=1$, $\gamma=10$, $\lambda=40$.}\label{fig:snapshot2d}
\end{figure}

We first consider a microscopic lattice gas which exhibit MIPS. \if{For the
sake of clarity, we detail only the one-dimensional case whose
extension to higher dimensions is straightforward.}\fi $N$ particles
evolve on a discrete ring of $\alpha L$ sites. There are two types of
particles and each site is occupied by at most one particle so that a
configuration can be represented using occupation numbers $\gs_i$ at
site $i$ with values in $\{-1,0,1\}$. To model self-propulsion, we
endow the $+$ particles with a weak drift to the right and the $-$
particles with a weak drift to the left, in addition to a symmetric
diffusive motion. Furthermore, a particle can tumble and change sign
at fixed rate. More precisely, the dynamics combine 3 mechanisms:
\begin{enumerate}
\item[1.1] For each bond $(i,i+1)$,  $\gs_i$ and
  $\gs_{i+1}$ are exchanged at rate D.
\item[1.2] For each bond $(i,i+1)$, a $+$ particle in $i$ jumps to
  $i+1$ if $\gs_{i+1} =0$ or a $-$ particle in $i+1$ jumps to $i$ if
  $\gs_i =0$, with rate $\gl/L$.
\item[1.3] Particles switch sign at rate $\gga/L^2$.
\end{enumerate}
The total number of particles is $N\equiv \gr_0 \ga L$ where $ \gr_0
\in [0,1]$ stands for the mean density. The system remains homogeneous
for small $\rho_0$ or $\lambda$, whereas the homogeneous phases become
unstable for large densities and drift.  The previous dynamics can be
generalized to higher dimensions.  We show in
Fig.~\ref{fig:snapshot2d} the result of 2D numerical simulations
leading to the coexistence between dilute and dense phases typical of
MIPS. Depending
on whether the 2D case is built solely with a left-right bias or
whether one considers biases along each of the four directions, we
observe different symmetries for the coexistence phases.

To account for this phenomenology, one needs to characterize the
evolution of the local density of particles. The expectation of the
microscopic variables can be computed from the dynamical rules. For
example, let $\gs_i^\pm (t) = 1_{\{ \gs_i (t) = \pm 1\}}$ then
\begin{align}
\label{eq: average evolution}
\partial_t  \bra \gs_i^+  \ket
= & D[\bra \gs_{i+1}^+  \ket + \bra \gs_{i-1}^+  \ket - 2 \bra \gs_i^+  \ket] - \frac{\gga}{L^2} \big[  \bra \gs_i^+  \ket - \bra \gs_i^-  \ket \big] \nonumber\\
& \!\!\!\!\!\! + \frac{\gl}{L} \big[  
 \big \bra \gs_{i-1}^+  \big( 1 - |\gs_i| \big) \big \ket
 - \big \bra \gs_i^+  \big( 1 - |\gs_{i+1}| \big) \big \ket
\big]
\end{align}
These equations are however not closed, since the evolution of $\bra
\gs_i^+(t)\ket$ involves the correlator $\bra\gs_i^+(t)|
\gs_{i+1}(t) | \ket$.

A closed, explicit description of the dynamics can, however, be
achieved at the macroscopic level. Indeed,
following~\cite{DeMasi:1985:PRL,JonaLasinio:Proba:1993,
  Bodineau:2010:JSP, Erignoux:2016:Arxiv}, we chose the three
processes above to occur with rates scaling with $L$ in such a way
that they all contribute equally to a hydrodynamic regime obtained by
a diffusive rescaling of space and time: $x=i/L$ and $\tau=t/L^2$.
Indeed, the first dynamical rule leads to the diffusion of the
particles: if one follows the particles without their signs, the
exchange dynamics 1.1 amount to a symmetric simple exclusion process
(SSEP). This first rule makes particles travel a distance $\Delta
i\sim L$ on a time $\delta t \sim L^2$ and hence at a macroscopic
scale $\Delta x\sim 1$ for $\Delta \tau \sim 1$. The second rule
applies at a reduced rate $\gl /L$, but provides a systematic drift to
the left or to the right depending on the particle type. Similarly, in
a time $L^2$, this leads to a displacement of order $L$. Finally at an
even more reduced rate $\gga/L^2$, the particle type changes which
boils down to saying that a particle changes direction once in a
macroscopic unit of time. This occurs sufficiently rarely so that the
drift has a macroscopic effect between two updates.

\begin{figure}
  \includegraphics[scale=1]{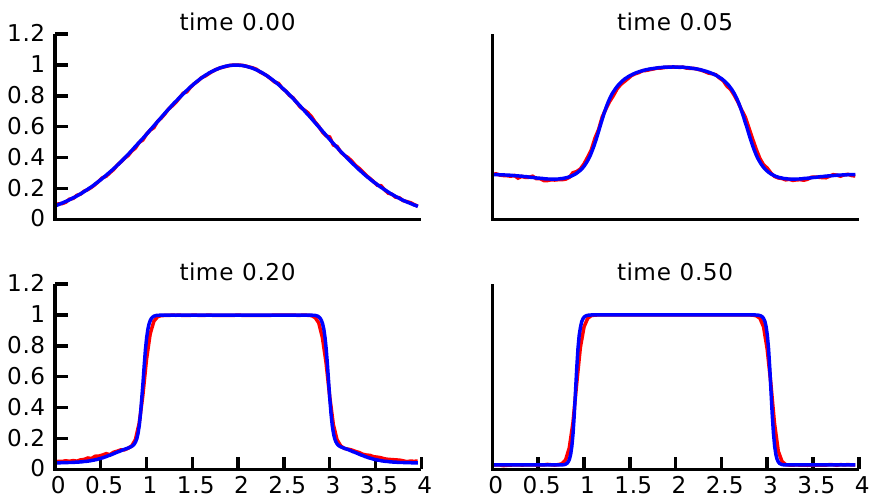}
 \caption{Successive snapshots leading to phase coexistence in 1D: the
   microscopic (red) and macroscopic (blue) simulations agree
   quantitatively. Simulation parameters: $D=1$, $\lambda=5$,
   $\gamma=0.1$, $\rho_0=0.75$, $\alpha=4$. {\bf Microscopic
     simulations:} Continuous time simulation of dynamics 1.1-1.3 with
   $L=1000$. The density profiles of the microscopic simulations are
   averaged over 200 independent runs; {\bf Macroscopic simulations:}
   Semi-spectral method with $n=50$ modes and a semi-implicit Euler
   time-stepping with $dt=10^{-4}$.} \label{fig:snapshot_time}
\end{figure}

To derive the hydrodynamic description of the system in the large $L$
limit, we introduce the macroscopic densities as
\begin{align}
  \gr^\pm (x, \tau) &\simeq \frac{1}{2 L^\gd} \sum_{|i - L x | \leq L^{\gd}} 1_{ \{ \gs_i ( \tau L^2) = \pm1 \} } , 
\label{eq: density}
\end{align}
where the coarse graining scale is determined by the parameter $\gd
\in (0,1)$.  Note that the microscopic system size $\alpha L$ depends
on two parameters: $1/L$ plays the role of a microscopic mesh; it
vanishes in the $L\to\infty$ limit in which $\alpha$ then controls the
rescaled system size $x\in [0,\alpha]$. The macroscopic equations for
the densities $(\gr^+,\gr^-)$, starting from a smooth initial
condition, can then be derived exactly as
\begin{align}
\label{eq: field rho+}
\partial_\tau \gr^+ &= D\partial^2_x  \gr^+ - \gl \partial_x [ \gr^+ (1- \gr) ] - \gga (\gr^+ - \gr^-),\\
\partial_\tau \gr^- &= D\partial^2_x   \gr^- + \gl \partial_x [ \gr^- (1- \gr) ] + \gga (\gr^+ - \gr^-),
\label{eq: field rho-}
\end{align}
where the total density is $\gr = \gr^+ + \gr^-$.  The mathematical
method to rigorously derive these hydrodynamic equations has been
initiated in~\cite{DeMasi:1985:PRL,JonaLasinio:Proba:1993} and we
refer to~\cite{Erignoux:2016:Arxiv} for a detailed implementation.  We
will explain below the underlying principles. Intuitively,
Eq.~\eqref{eq: field rho+} can be deduced from the microscopic
equation~\eqref{eq: average evolution} by first replacing the discrete
differences by derivatives. Justifying the forms of the non-linear
advection terms require to close the two-point correlations
in~\eqref{eq: average evolution}. Even though on a macroscopic scale
the three mechanisms of the dynamics compete at equal footing, the
first one dominates locally as it occurs much more frequently. Thus,
in the large $L$ limit, it can be shown that the local correlations
are controlled by the stirring part.  The invariant measures of the
dynamics reduced to the stirring part are product Bernoulli measures
indexed by two parameters which prescribe the local densities of $\pm$
particles.  Thus, at any time, the local statistics of the full
dynamics are determined by a product of Bernoulli measures
parametrized by the local densities \eqref{eq: density}.  The
approximation by these local measures is valid beyond the expectation
of the correlations and applies at the level of sample paths so that
local averages as in \eqref{eq: density} converge with high
probability to the solution of the hydrodynamic equations \eqref{eq:
  field rho+}--\eqref{eq: field rho-}. Note that the hydrodynamic
equations directly extend to higher dimensions. We compare in
Fig.~\ref{fig:snapshot_time} simulations of the microscopic and
macroscopic dynamics for the 1D case. A perfect agreement between the
two dynamics is observed on their way to phase coexistence.

To analyze the emerging behaviours predicted by~\eqref{eq: field
  rho+}-~\eqref{eq: field rho-}, we first introduce an unnormalized
polarization field $m = \gr^+ - \gr^-$. The dynamics can then be
reduced to a dimensionless form using $\rho = \bar \rho \tilde\rho$, $
m = \bar \rho \tilde m $, $ x = \ell \tilde x $ and $ t = \tau \tilde
t$ where $\bar \rho = 1$, $\tau = \frac{1}{\gamma}$, $\ell =
\sqrt{\frac{D}{\gamma}}$, so that $\tilde x\in [0;\alpha
  \sqrt{\frac{\gamma}{D}}]$. In this system of units the evolution
equation reads:
\begin{align}
  \label{eq:mips_adim_rho}
 \partial_{t} \rho &= \Delta \rho   - \Pe \nabla (m(1-\rho))\\
  \label{eq:mips_adim_m}
 \partial_{t} m &= \Delta m   -  \Pe \nabla (\rho(1-\rho)) - 2 m
\end{align}
with $\Pe=\frac{\lambda}{\sqrt{D\gamma}}$ and where, for the sake of
clarity, we have omitted the tilde notation (we stick to the rescaled
variables until the end of the discussion of this model).
Eqs.~\eqref{eq:mips_adim_rho} and~\eqref{eq:mips_adim_m} show that the
system is fully characterized by two control parameters: the density
$\rho_0=N/L$ and the Péclet number \Pe. The latter compares the
length traveled between two tumbles thanks to the drift,
${\lambda}/{\gamma}$, to the one resulting from the diffusive dynamics,
$\sqrt{{D}/{\gamma}}$.   For small P\'eclet numbers, the diffusion dominates and the effect of
self-propulsion is negligible. Conversely, the effect of activity gets
more and more pronounced as \Pe{} increases.

The homogeneous solutions of
Eqs.~\eqref{eq:mips_adim_rho}--\eqref{eq:mips_adim_m}, $\rho(x,t) =
\rho_0$ and $m(x,t)=0$, are linearly unstable when
\begin{equation}
\Pe^2(1-\rho_0)(2\rho_0-1)> 2.
\end{equation}
For any \Pe{} larger than a critical value ${\rm Pe^c}=4$, the system
is thus linearly unstable for $\rho_0 \in [\rho^s_l,\rho^s_h]$ with
$\rho^s_{l,h} = \frac{3}{4} \pm
\frac{1}{4}\sqrt{1-\frac{16}{\Pe}}$. This defines the spinodal region
of the system. Note that this is a large wavelength instability,
observed only for macroscopic system size $L_\alpha\equiv \alpha\sqrt{\frac{\gamma}{D}} >
\frac{2 \pi}{\sqrt{\Pe^2(1-\rho_0)(2\rho_0-1)-2}}$. 
We now turn to the computation of the coexisting densities, generalizing the method
introduced in~\cite{Solon:2016:Arxiv}.

\begin{figure}
  \includegraphics[width=\columnwidth]{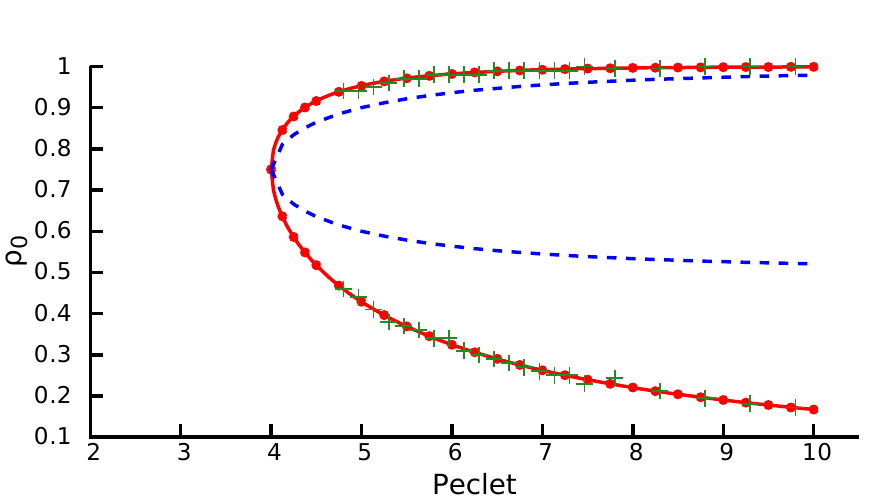}
  \caption{ \label{fig:phase_diagram_mips} Phase diagram of the MIPS
    observed for dynamics 1.1-1.3. For each Peclet number, the
    coexisting phases are computed analytically (red line), by
    simulating the microscopic process (black \& green dots) and by
    numerically solving the macroscopic equations (red crosses). The
    analytic predictions for the spinodals are shown in blue.}
\end{figure}

For simplicity, we consider the 2D case with left-right bias in which
the interfaces between the phases are flat and along $\hat y$. We
consider fully phase-separated profiles and use
Eqs.~\eqref{eq:mips_adim_rho}--\eqref{eq:mips_adim_m} to construct a
domain-wall solution describing the evolution of the density and
magnetization fields through an interface. In the steady-state,
Eq.~(\ref{eq:mips_adim_rho}) simply leads to $m=\frac{1}{\Pe}
\frac{\nabla \rho}{1-\rho}$. Eq.~(\ref{eq:mips_adim_m}) then reads
$\partial_x g=0$ with
\begin{align}
g \equiv g_0(\rho) + \Lambda(\rho) (\partial_x \rho)^2 - \kappa(\rho) \partial_{xx} \rho 
\end{align}
where $\Lambda(\rho)^{-1} = -{\Pe(1-\rho)^2}$, $\kappa(\rho)^{-1} =
{\Pe(1-\rho)}$, and $g_0(\rho)= \Pe \,\rho(1-\rho) - {2}
\log(1-\rho)/{\Pe}$. Since the density is homogeneous in the gas and
liquid phases, one gets a first relationship between the coexisting
densities:
\begin{equation}
  \label{eq:contition_mips_1_g}
  g_0(\rho_g) = g_0(\rho_\ell) \equiv \bar g(\Pe)
\end{equation}
We now introduce a function $R(\rho)$
such that $R'' \kappa= -(2 \Lambda+\kappa') R'$ and $\phi(R)$ such
that $\phi'(R)=g_0(\rho)$. Computing $I=\int_{x_g}^{x_\ell} g
\partial_x R dx$ leads to
\begin{eqnarray} \label{eq:integralfreecountry}
  I= \bar g [R(\rho_\ell)-R(\rho_g)] = \Phi(R_\ell) - \Phi(R_g)\;.
\end{eqnarray}
Eq.~\eqref{eq:integralfreecountry} then enforces the equality of
$h_0\equiv\phi'(R) R-\phi(R)$ between the two phases. The function $R$
and $\phi$ can be computed explicitly as
\begin{equation}
R(\rho) = \log(1-\rho);\quad \Phi(R)=\Pe(1-\frac {e^R}2)e^R-\frac{R^2}\Pe
\end{equation}
The binodals can then be computed from the equality of $g_0$ and $h_0$
between the two phases, which amounts to a common tangent construction
on $\phi(R)$. The resulting phase diagram is shown in
Fig~\ref{fig:phase_diagram_mips}. It shows perfect agreement with both
simulations of the hydrodynamic equations and of the microscopic
models in 1D and 2D. As far as we are aware, this is the first
microscopic model for which the hydrodynamic description and the
phase diagram of a motility-induced phase separation can be
derived exactly.

We now turn to the phase transition leading to collective motion which
is probably the most studied emerging behaviour in active
matter~\cite{Vicsek:1995:PRL,Toner:1995:PRL,Gregoire:2004:PRL,Solon:2013:PRL,Solon:2015:PRL,Marchetti:2013:RMP,Schaller:2010:Nature,Bricard:2013:Nature}. Following
the strategy laid out in the first part of this letter, we introduce a
microscopic model of polar aligning active particles and derive its
hydrodynamic limit exactly. For simplicity, we first describe the
model in one dimension. $N$ particles evolve on a discrete ring of
$\alpha L$ sites. Each particle is described by two degrees of
freedom: its position $i\in\{1\dots L\}$ and its orientation, noted
$\pm$ in 1D. We call $\eta_i = (\eta_i^+, \eta_i^-) \in \N^2$ the
number of particles of each type on site $i$. The dynamics of a
$\pm$ particle is given by the three following processes:
\begin{enumerate}
 \item[2.1] Symmetric hops with rate $2D$
 \item[2.2] Jumps from site $i$ to $i \pm 1$  with rate $\frac{\lambda}{L}$
 \item[2.3] Flips into a $\mp$ particle  with rate $\frac{1}{L^2}c^\pm_i(\eta_i^+,\eta_i^-)$
\end{enumerate}
where we choose $c^\pm$ to produce a polar alignement
\begin{align}
  c^\pm(\eta_i^+,\eta_i^-) = \exp[\mp \gb(\eta_i^+-\eta_i^-)]
\end{align}

\begin{figure}
  \includegraphics[width=\columnwidth]{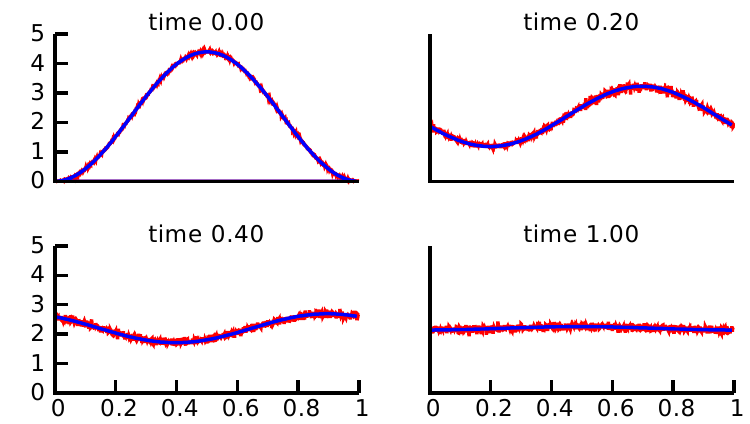}
  \caption{Successive snapshots of one-dimensional simulations of the
    microscopic dynamics 2.1-2.3 and of the hydrodynamic
    equations~\eqref{eq:hydrorho} and~\eqref{eq:hydrom}. The system is
    in the ordered phase and the initial condition is fully ordered
    with $\rho(x,t)=\rho_0[1+\cos(2\pi x)]$. {\bf Microscopic
      simulations:} the density profiles are obtained by averaging
    over 300 continuous time simulations of the dynamics 2.1-2.3 on a
    lattice of $L=1000$ sites ($\alpha=1$); {\bf Macroscopic
      simulations:} pseudo-spectral simulations with 50 Fourier modes
    and a semi-implicit time-stepping with $dt=10^{-4}$. Physical
    parameters: $D=0.5,\,\lambda=4,\beta=0.8$.
  }\label{fig:Isingrelax}
\end{figure}

We consider again a diffusive rescaling of time and space to obtain
the exact hydrodynamic equations
\begin{align}
  \label{eq:rhoplusactiveising}
  \partial_t \gr^+ &=D \gD \gr^+ + \gl \nabla \gr^+ - F(\gr^+, \gr^-) \\ 
  \label{eq:rhominusactiveising}
  \partial_t \gr^- &= D \gD \gr^- - \gl \nabla \gr^- + F(\gr^+, \gr^-)
\end{align}
where the functions $F(\gr^+,\gr^-)$ are given by
\begin{equation}
  F(\rho^+,\rho^-)=f^+(\rho^+,\rho^-)-f^-(\rho^+,\rho^-)
\end{equation}
and $f^\pm(\rho^+(x),\rho^-(x))$ are the averages of $n^{ \pm}
c^\pm(n_i^+,n_i^-)$  with respect to the local Poisson measure
\begin{equation}
  \nu_{\rho^+,\rho^-}(n_i^+,n_i^-)={\rm e}^{-\rho^+-\rho^-}\frac{(\rho^+)^{n_i^+}}{(n_i^+)!}\frac{(\rho^-)^{n_i^-}}{(n_i^-)!}\;.
 \label{eq: mesure Poisson}
\end{equation}
Again, while the dynamical rules $2.1-2.3$ all contribute equally in
the hydrodynamic scaling, the symmetric random walk equilibrates much
faster on the microscopic mesh scale. The averages of the non-linear
contributions due to the flipping rules are thus computed with respect
to local Poisson measures, which are the steady-state measures of the
symmetric random walk, conditionned to producing the correct mean
local densities of $+$ and $-$ particles. Finally, the hydrodynamic
equations~\eqref{eq:rhoplusactiveising}--\eqref{eq:rhominusactiveising}
can be rewritten in the more familiar form
\begin{align}
  \label{eq:hydrorho}
  \partial_t \rho &= D \Delta \rho + \lambda \nabla m\\ 
  \label{eq:hydrom}
  \partial_t m &= D \Delta m + \lambda \nabla \rho - 2\tilde F(m,\rho) 
\end{align}
where 
\begin{align}
  \tilde F = \big(
    m\, \cosh[m\, \sinh(\beta)]      -\rho\, \sinh[m\, \sinh(\beta)]\big) e^{-\beta+\rho \,\cosh(\beta)-\rho}
\end{align}
Note that $\tilde F$ is \textit{not} equal to the mean-field
expectation of $n^+c^+-n^-c^-$.

Simulations of the microscopic model and its hydrodynamic description
confirm the presence of a transition to collective motion. At large
`temperature' $T\equiv\beta^{-1}$ and low density $\rho_0=N/V$, the
system is in a homogeneous disordered `gas' phase. At low noise and
large density, the system is in a homogeneous ordered liquid phase
with a non-zero average flux of particles in the steady state. These
homogeneous phases are separated by a coexistence region in a which a
dilute disordered gas of density $\rho_g(T)$ coexist with a dense
liquid phase of density $\rho_\ell(T)$. A perfect agreement between
simulations of the microscopic model and of its hydrodynamic
description is shown in Fig.~\ref{fig:Isingrelax} where the relaxation
of a perturbation towards a homogeneous liquid phase is shown.
\begin{figure}
  \hspace{-1cm} \includegraphics[width=.9\columnwidth]{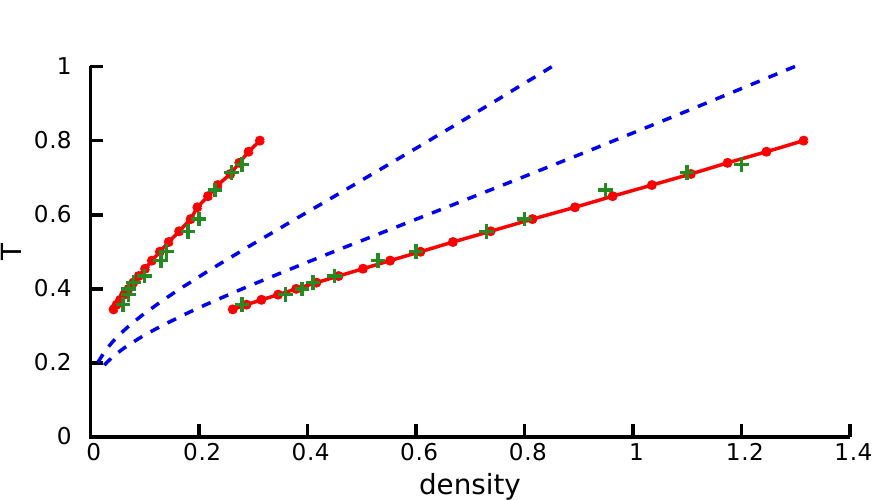}
  \caption{Phase diagram of the flocking transition observed for
    dynamics 2.1-2.3 in the $T,\rho$ plane for $D=0.5$ and
    $\lambda=1$. The spinodals are shown in blue. Simulations of the
    hydrodynamics equations (red lines) and the microscopic model
    (green dots) lead to coexisting densities which are in agreement
    within the accuracy of our simulations.
  }\label{fig:phase_diagram_flock}
\end{figure}

For this microscopic model, the similarity of this transition to
collective motion with a liquid-gas phase transition in the canonical
ensemble can now be derived analytically. The hydrodynamic description
indeed predicts that the disordered homogeneous phase
$\rho(x)=\rho_0,m(x)=0$ loses linear stability for densities such that
$\partial_m \tilde F(0,\rho) <0$, i.e. $\rho_0 > \rho_s^{g}\equiv
\sinh(\beta)^{-1}$. Then, a fully ordered solution appears given by
$\rho(x)=\rho_0$ and $m_0$ the solution of
\begin{align}
  \label{eq:mzeroequation}
\frac{m_0}{\rho_0} - \tanh[m_0\sinh(\beta)] = 0\;.
\end{align}
This solution is however linearly unstable for
$\rho_s^g<\rho_0<\rho_s^\ell$ and leads to the travelling bands
described above. As far as we are aware, this is the first time that the
existence of a region in which neither homogeneous phases is stable is
rigorously proven. At higher densities, for $\rho_0\geq \rho_s^\ell$, the
ordered uniform state becomes linearly stable. In agreement with the
liquid-gas scenario proposed for the flocking
transition~\cite{Solon:2013:PRL,Solon:2015:PRL}, the coexisting
densities are such that $\rho_g< \rho_s^g<\rho_s^\ell<\rho_\ell$. The
corresponding phase diagram is shown in
Figure~\ref{fig:phase_diagram_flock}. The coexisting densities
observed for the microscopic model agree with their macroscopic
counterparts within the numerical accuracy of our simulations. 

In this Letter we have introduced a class of lattice models for which
one can derive exact hydrodynamic equations. Our strategy relies on
scaling the rates of the different dynamical contributions so that
they contribute equally at a diffusive hydrodynamic scale. The
symmetric hopping however controls the dynamics at a local, mesoscopic
scale. This yields an explicit form for the local measure one needs to
use to compute the average of any non-linear function entering the
dynamics of the mean local densities. Then, the hydrodynamic
descriptions allow us to characterize the large scale emerging
behaviours of these active lattice gases. In particular we have
introduced two models presenting two of the most studied collective
behaviours of active particles, namely the Motility-Induced Phase
Separation and the transition to collective motion.  Constructing more
general models is rather straightforward using the ingredients
presented in this letter, for instance to study nematic alignement or
the interplay between MIPS and aligning torques. {Note that the
  hydrodynamic description is exact for finite macroscopic times in
  the $L\to\infty$ limit. For large-but-finite sizes, describing the
  statistics of the active lattice-gas trajectories requires the
  addition of sub-leading fluctuating terms, in the spirit of the
  Macroscopic Fluctuation
  Theory~\cite{Bertini2002,Derrida:2007:JSM,Tailleur2008JPA}. These
  terms are key to understanding the selection of meta-stable
  propagating solutions observed in simple flocking
  models~\cite{Caussin2014PRL,Solon2015PRE,Solon:2015:PRL}; their
  rigorous mathematical derivation, however, remains an open
  challenge.}

\textit{Acknowledgment} We thank Alexandre Solon and Hugues Chaté for
interesting discussions. JT and MKH acknowledge support from the ANR
grant Bactterns. CE and TB acknowledge the support of
ANR-15-CE40-0020-01 grant LSD.  C.E. is supported by the Brasilian
National Council for Scientific and Technological Development
(CNPq). Part of this work was done during the author's stay at the
Institut Henri Poincare - Centre Emile Borel during the trimester «
Stochastic Dynamics Out of Equilibrium ». The author thanks this
institution for its support.

\bibliographystyle{apsrev4-1}
\bibliography{biblio}
\end{document}